\documentclass[%
 reprint,
 amsmath,amssymb,
 aps,
]{revtex4-1}

\usepackage{hyperref}
\usepackage{graphicx}% Include figure files
\usepackage{dcolumn}% Align table columns on decimal point
\usepackage{bm}% bold math
\usepackage[caption=false]{subfig}

\begin{document}

\preprint{APS/123-QED}

\title{Comment on ``Replica Symmetry Breaking in Trajectories of a Driven Brownian Particle"}

\author{Tapas Singha} 
\email{tapass@tifrh.res.in}
\author{Mustansir Barma}
\email{barma23@gmail.com}
\affiliation{TIFR Centre for Interdisciplinary Sciences, Tata Institute of Fundamental Research, Gopanpally, Hyderabad 500107, India}

\date{\today}% It is always \today, today,

\maketitle

 Recently Ueda and Sasa (US) \cite{Ueda_Sasa2015} studied the overlap of the trajectories of two passive particles subject to driving by the same noisy Burgers field \cite{KPZ_1986}, but independent noise. A quantity of central importance is the probability distribution $P_t (d)$ of the interparticle distance $d$ at time $t$, whose behavior indicates an interesting coexistence of a high-overlap and low-overlap phase in space-time \cite{Ueda_Sasa2015}. Contrary to the stretched exponential form reported in \cite{Ueda_Sasa2015}, in this Comment we show that $P_t(d)$ in fact follows a power law, and give strong numerical evidence for this. A crucial point is that $P_t(d)$ is a \emph{scaling function} of $d$ and $t$. Besides its theoretical importance, the scaling form immediately explains 
 US's numerical findings: (i) the linear (in time) growth of mean-squared separation (ii)
 the surprising time independence of probability density. It also explains why the overlap 
 vanishes for Edwards-Wilkinson (EW) driving in the large-time limit.

\begin{figure}[ht!]
\centering
\begin{minipage}{0.48\textwidth}
\includegraphics[width=\textwidth,height=.21\textheight]{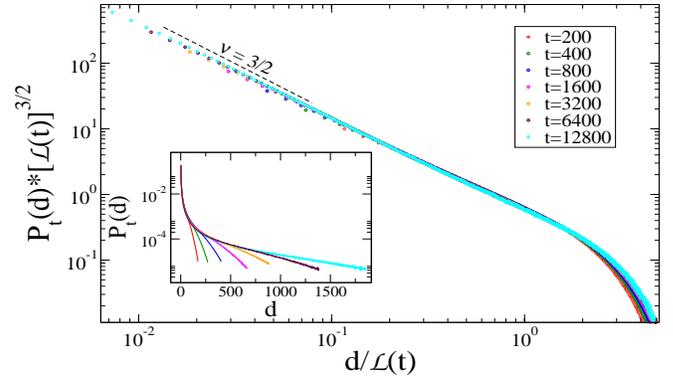}
%\caption{first}
\end{minipage}%
\caption{The probability distribution $P_t(d)$ with KPZ driving shows a scaling collapse. Inset: same data in a semi-log plot.}
\label{Fig_PDF_KPZ}
\end{figure}

A recent study of the approach to steady state of passive sliders on a Kardar-Parisi-Zhang (KPZ) surface revealed an indefinitely growing coarsening length scale $\mathcal{L}(t) \sim t^{1/z}$, underlying a scaling description \cite{Singha_Barma2018}. Here $z=\frac{3}{2}$ is the KPZ dynamic exponent. For the problem considered in \cite{Ueda_Sasa2015}, the initial condition is one in which all particles start at the same location. We checked that scaling remains valid, through numerical simulations of particles on a stochastically evolving lattice with KPZ characteristics \cite{Singha_Barma2018} as shown in Fig.\ \ref{Fig_PDF_KPZ}.  In the limit of large $d$ and $t$ with $d/t^{1/z}$ held fixed, Fig.\ \ref{Fig_PDF_KPZ} shows that $P_t(d)$ assumes the scaling form  
\begin{equation}
P_t(d) \approx  \frac{1}{[\mathcal{L}(t)]^{(1+\theta)}} \  Y \left( \frac{d}{\mathcal{L}(t)} \right)
\label{KPZ_PDF}
\end{equation}
where $Y(y)\sim y^{-\nu}$ as $y \rightarrow 0$, and falls exponentially as $y \rightarrow \infty$. We find $\theta \simeq \frac{1}{2}$ and $\nu \simeq \frac{3}{2}$ consistent with steady state values, where $\mathcal{L}(t)$ in Eq.\ \ref{KPZ_PDF} plays the role of system size \cite{Apurba2006}. The scaling form leads to a convincing data collapse, in contrast to the unscaled semi-log plot (inset of Fig.\ \ref{Fig_PDF_KPZ};  see also Fig.\ 3 of Ref. \cite{Ueda_Sasa2015}) which was analyzed in terms of a stretched exponential form \cite{Ueda_Sasa2015}.

 The scaling form Eq.\ \ref{KPZ_PDF} has important implications : 
(i) For large $d$ and $t$, the mean-squared separation $\langle d^2\rangle$ is straightforwardly seen to obey $\langle d^2 \rangle \sim t$ as found numerically in \cite{Ueda_Sasa2015}. (ii) For large values of $t$, Eq.\ \ref{KPZ_PDF} implies that $P_t(d)$ becomes time independent for $d < d^{*}$, where $d^{*} \sim  \mathcal{L}(t)$. This explains the observation in \cite{Ueda_Sasa2015} that curves for $P_t (d)$ coincide for $d <60$ in the time range $200 \leq t \leq 1000$. Importantly, the fact that $d^{*}$ increases with $t$ implies that trajectory pairs would show overlap (Eq.\ 3 of \cite{Ueda_Sasa2015}) with even larger values of localization length. Finally, the scaling form (1) holds also for the case of EW driving, but with $\theta \simeq 0$, $\nu \simeq \frac{2}{3}$ and $z=2$ with logarithmic corrections \cite{Huveneers}. Thus for all $d$, the probability $P_t(d)$ falls to zero   as $ [\mathcal{L}(t)]^{-1/3} \sim t^{-\frac{1}{6}}$, in contrast to the KPZ case. This explains why the overlap approaches zero as $t \rightarrow \infty$ in the EW case \cite{Ueda_Sasa2015}.

\vskip 0.5cm

\end{document}